\documentclass[rnote]{aa}

\usepackage{txfonts}
\usepackage{natbib}
\bibpunct{(}{)}{;}{a}{}{,}
\usepackage{graphicx}
\usepackage {epstopdf}

\begin{document}

\title{No apparent accretion mode changes detected in Cen\,X-3}

\author{Daniela M\"uller \inst{\ref{inst1}}
\and Dmitry Klochkov \inst{\ref{inst1}}
\and Andrea Santangelo \inst{\ref{inst1}}
\and Tatehiro Mihara\inst{\ref{inst2}}
\and Mutsumi Sugizaki\inst{\ref{inst2}}
}
\institute{Institut f\"ur Astronomie und Astrophysik, Universit\"at T\"ubingen, Sand 1, D-72076 T\"ubingen, Germany,\\
e-mail: daniela.mueller@astro.uni-tuebingen.de\label{inst1} \and
MAXI team, RIKEN, 2-1 Hirosawa, Wako, Saitama 351-0198\label{inst2}}
\date{Received /
Accepted}

\abstract {} 
{The presence of two distinct spectral states has previously been reported 
for Cen\,X-3 on the basis of RXTE/ASM observations. Triggered by this 
result, we investigated the spectral properties of the source using the
larger amount of X-ray data now available with the aim to clarify 
and interpret the reported behavior.}
{To check the reported results we used the same data set and followed the
same analysis procedures as in the work reporting the two spectral states.
Additionally, we repeated the analysis using the enlarged data sample 
including the newest RXTE/ASM observations as well as the data from the 
MAXI monitor and from the INTEGRAL/JEM-X and ISGRI instruments.} 
{We could not confirm the reported presence of the two spectral states in 
Cen\,X-3 either in the RXTE/ASM data or in the MAXI or INTEGRAL data. 
Our analysis showed that the flux variations in different energy bands are
consistent with the spectral hardness being constant over the entire time 
covered by observations.
}
{}

\maketitle

\section{Introduction}
 
Cen\,X-3 was first observed in 1967 \citep{1967PhRvL..19..681C} and later 
identified as a pulsating high-mass X-ray binary system 
based on the \emph{Uhuru} observations 
\citep{1971ApJ...167L..67G, 1972ApJ...172L..79S}. 
The pulsation period of the source was found to be 
$\sim$4.8\,s \citep{1971ApJ...167L..67G},
the orbital period (determined from regular X-ray eclipses) is
$\sim$2.08\,days \citep{1972ApJ...172L..79S}. Measuring the eclipse times, \citet{1974ApJ...192L.139A} derived a mass of 0.6--1.1\,$M_{\sun}$ for the compact 
object and of 16.5--18.5\,$M_{\sun}$ for the companion star, well in 
accordance with \citet{1979ApJ...229.1079H} who determined the masses from the
optical spectroscopic observations of the companion. The projected orbital 
radius was determined from the analysis of pulse arrival times to be 
$\sim$39.75$\pm$0.04\,light-seconds \citep{1972ApJ...172L..79S}. 
The system is located at a distance of about 8\,kpc 
\citep{1974ApJ...192L.135K}, with a lower limit of 6.2\,kpc 
\citep{1974ApJ...192L.135K}. Cen\,X-3 shows a non-periodic alternation of
high and low states with a characteristic time of reccurence of 
125--165\,days, derived from analysis of \emph{Vela} data 
\citep{1983ApJ...273..709P}. This long-term variability is sometimes attributed
to a precessing accretion disk \citep{1983ApJ...273..709P}.\\

Our analysis was triggered by the work of \citet{2005A&A...442L..15P}. 
Based on their analysis of RXTE/ASM data of Cen\,X-3, the authors reported
the presence of two distinct spectral states/modes in the high state of the 
source distinguished by the hardness ratio. In the high state the peak flux is up to a factor of 40 larger than during the low state \citep{2005A&A...442L..15P}. It was found that when the source makes a transition from a low to a 
high state, it adopts one of these two spectral modes and during the entire
high-intensity phase remains in that mode. \citet{2005A&A...442L..15P}
showed that during all high states between December 2000 and 
April 2004 the source was in the hard spectral mode as indicated by large
hardness ratio, while in all high states prior and subsequent to this period
it was in the soft spectral mode. To exclude systematic effects the authors analyzed ASM data on three other sources:
Her\,X-1, Vela\,X-1 and SMC\,X-1. But none of the sources showed the behavior
similar to that reported for Cen\,X-3. \citet{2005A&A...442L..15P} interpreted
their results as a manifestation of two accretion modes which are at work
in Cen\,X-3 at different times.

We repeated the analysis of \citet{2005A&A...442L..15P}
using the same data set and following the analysis procedures described
in their paper. We then extended our analysis to the entire ASM data on
the source available by now and included the data obtained with the
MAXI, INTEGRAL/JEM-X and INTEGRAL/ISGRI instruments.

\section{Observational data and analysis method}

We used data from the All Sky Monitor (ASM) onboard the Rossi X-ray Timing
Explorer (RXTE), from the Monitor of All Sky X-ray Image (MAXI) and from the JEM-X and ISGRI instruments of the INTEGRAL (INTErnational Gamma-Ray Astrophysics Laboratory) satellite.\par\medskip

\begin{figure*} \centering
\includegraphics[viewport=1cm 0cm 13.5cm 25.5cm,scale=0.72,clip,angle=270]{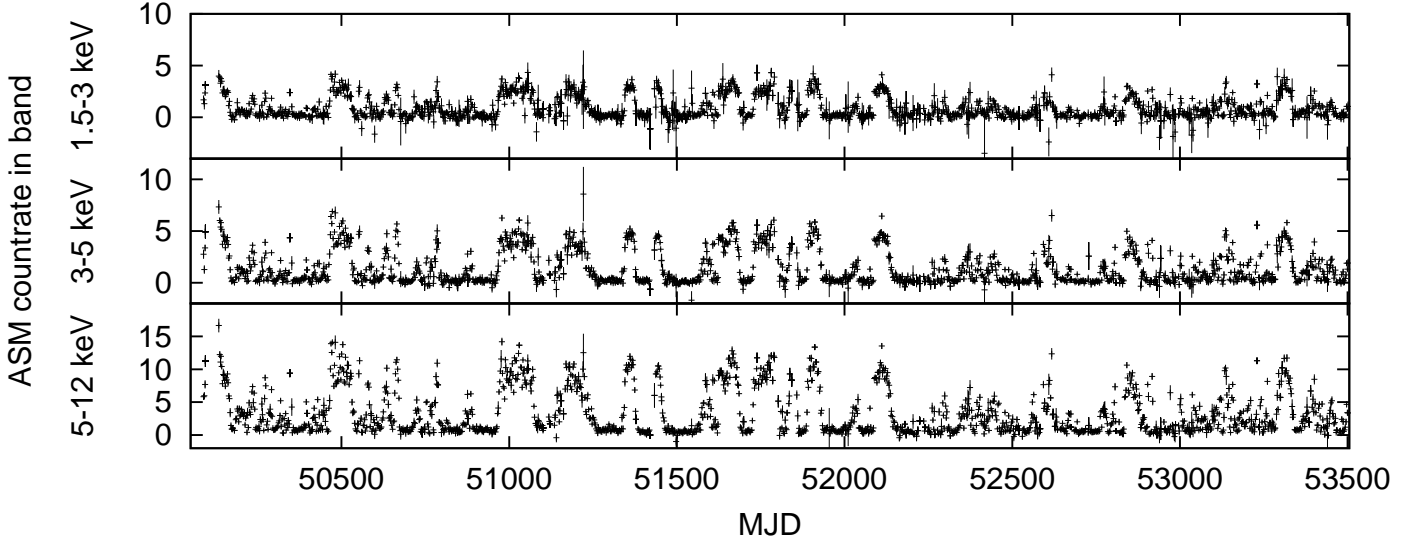}
\caption{The processed ASM lightcurve of Cen\,X-3 from MJD 50087 to 53501. 
The three energy bands A (1.5--3\,keV), B (3--5\,keV) and C (5--12\,keV) 
are shown in different panels.}
\label{figure:1}
\end{figure*}

The ASM instrument is an X-ray monitor covering the 1.5--12\,keV energy 
range \citep{1996ApJ...469L..33L}. It consists of three Scanning Shadow 
Cameras (SSCs) each with a position-sensitive proportional counter 
\citep{1996ApJ...469L..33L}. 
The counts detected in three energy bands, 1.5--3, 3--5 and 5--12\,keV, 
are accumulated in dwells of about 90\,s. 
In our analysis we used the \textit{Dwell by Dwell} data available from 
the ASM team on the web (\textit{http://xte.mit.edu/asmlc/}). We checked that these data differ only in format from the data available at the NASA's High Energy Astrophysics Science Archive Center (HEASARC) which were used by \citet{2005A&A...442L..15P}. \par\medskip

MAXI is an X-ray monitor onboard of the International Space Station (ISS) 
\citep{2009PASJ...61..999M}. It scans the sky each orbit of 92\,
minutes observing a particular source for about 40--150\,seconds 
\citep{2011arXiv1102.0891S}, depending on the source position. 
MAXI has two cameras, the Gas Slit Camera (GSC) and the Solid-state Slit 
Camera (SSC). The main instrument, GSC, consists of proportional counters 
with slit and slat collimators \citep{2009PASJ...61..999M} 
with a total effective area of 5350\,$\mathrm{cm}^2$ 
\citep{2009PASJ...61..999M}. Every orbit, about 85\% of the sky is 
scanned in the 2--30\,keV energy band \citep{2011arXiv1102.0891S}. 
MAXI lightcurves with a time resolution of one orbit are available on 
the web (\textit{http://maxi.riken.jp}). The lightcurves are produced for 
three energy bands, 2--4\,keV, 4--10\,keV, and 10--20\,keV.\par\medskip

The INTEGRAL satellite observes objects simultaneously in gamma rays, X-rays and visible light. JEM-X is one of its instruments and consists of two telescopes with coded aperture masks \citep{2003A&A...411L.231L}. JEM-X obtains X-ray spectra and imaging in the 3--35\,keV energy band \citep{2003A&A...411L.231L}. IBIS is the gamma-ray imager of INTEGRAL and consists of two detector layers, ISGRI and PICsIT \citep{2003A&A...411L.131U}. ISGRI is the soft gamma ray imager \citep{2003A&A...411L.141L}. It consists of a CdTe detector camera with a sensitive area of 2621\,$\mathrm{cm}^2$ observing in the 15\,keV--1\,MeV energy band \citep{2003A&A...411L.141L}. INTEGRAL data can be downloaded from the web (\textit{http://www.isdc.unige.ch/heavens\_webapp/integral/}) for all instruments and for a number of predefined energy bands. For our analysis we selected following energy bands: JEM-X 3.0--5.5\,keV, 5.5--10.2\,keV, 10.2--18.9\,keV, 18.9--34.9\,keV, ISGRI 22.1--30.0\,keV, 30.0--40.3\,keV, 40.3--51.2\,keV, 51.3--63.3\,keV. \par\medskip

As mentioned above, our analysis follows the procedures described in 
\citet{2005A&A...442L..15P}. First, we excluded data falling in the intervals
of X-ray eclipses. We then averaged the data within each orbit of the system
to avoid
any spectral dependence on the orbital phase \citep{2008ApJ...675.1487S},
i.e. we produced light curves with one time bin per orbit. 
For consistency, we used the same orbital parameters as 
\citet{2005A&A...442L..15P}: the orbital period $P_{\rm orb}=2.08702$\,d,
the mid-eclipse time $t_{\rm mid}\,{\rm (TJD)} = 10087.295$, and the eclipse 
duration in units of orbital phase $\Delta\phi = 0.306$ (including ingress and egress).
First, we performed the analysis on the same ASM data set as used in 
\citet{2005A&A...442L..15P}: from MJD 50087 to 53501. Then, to enlarge the 
data sample, we added ASM data on the source covering the time range from 
MJD 53501 to 55646 (i.e. all ASM observations available by the time of 
preparation of this work). For the MAXI analysis we used all the data 
available by now, from MJD 55097 to MJD 55630. For the INTEGRAL analysis we used data from MJD 52650 to 54959 which is all available data in the archive.

\section{Results}
\subsection{RXTE/ASM results}
\begin{figure} [!h] \centering
\includegraphics[viewport=0.0cm 0cm 18.5cm 25cm,scale=0.33,clip,angle=270]{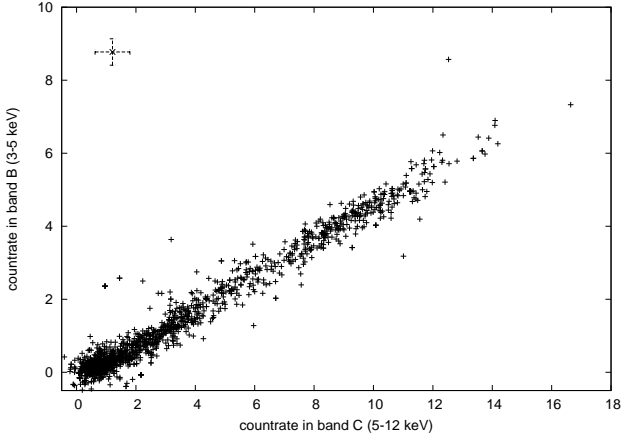}
\caption{ASM data between MJD 50087 and 53501 (same as used in 
\citealt{2005A&A...442L..15P}). Countrate in Band B (3--5\,keV) is plotted versus 
countrate in Band C (5--12\,keV). Typical uncertainties of the data
points are indicated by the error bars in the upper left corner of the plot.}
\label{figure:2}
\end{figure}

The resulting ASM lightcurve of Cen\,X-3
in the time range MJD 50087--53501
in three energy bands with removed 
eclipse intervals with each bin corresponding to one orbit of the system 
is shown in Figure\,\ref{figure:1}. In the 5--12\,keV light curve, one can 
already see substantial differences with respect to the results of 
\citet{2005A&A...442L..15P} (see Fig.\,1 in their paper).
The reported spectral hardening accompanied with an 
increase of the flux between MJD 51800 and 53100 in the highest energy band 
could not be confirmed in our analysis. 
The reported two spectral states should emerge as two branches of 
data points in a plot of the countrates in the 3--5\,keV band versus 
the 5--12\,keV band, as shown in Figure\,3 of 
\citet{2005A&A...442L..15P}. In the corresponding plot resulted from 
our (re-)analysis of the data (Fig.\,\ref{figure:2}), we could not 
find any hint of different branches.

\begin{figure} [!h] \centering
\includegraphics[viewport=0.0cm 0cm 18.5cm 25cm,scale=0.34,clip,angle=270]{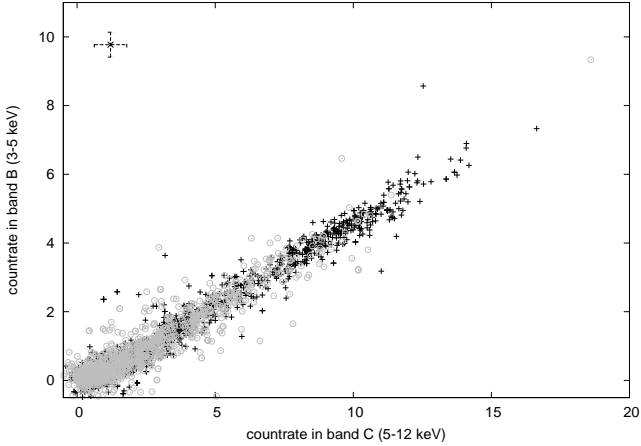}
\caption{ASM data: Countrate in band B (3--5\,keV) versus countrate in band C (5--12\,keV). The data from MJD 50087 to 53501 are marked with a cross to be distinguished from the data between MJD 53501 to 55646 which are marked by circles. Typical uncertainties of the data points are indicated by the error bars in the upper left corner of the plot.}
\label{figure:3}
\end{figure}

The inclusion of the new ASM data that became available since the work of
\citet{2005A&A...442L..15P} (MJD 53501 to 55646) also did not reveal any 
separation of the data points in two spectral states.
The two modes do not appear either in the new data alone or in the 
entire set of ASM data. Figure\,\ref{figure:3} shows the ASM fluxes in the 
3--5\,keV band versus fluxes in the 5--12\,keV band. The data from the 
new observations (after MJD 53501) are plotted with a different marker. 
We could not detect any change with respect to the older ASM observations.

\subsection{MAXI results}

\begin{figure} [!h] \centering
\includegraphics[viewport=0.0cm 8cm 17.5cm 22cm,scale=0.49,clip]{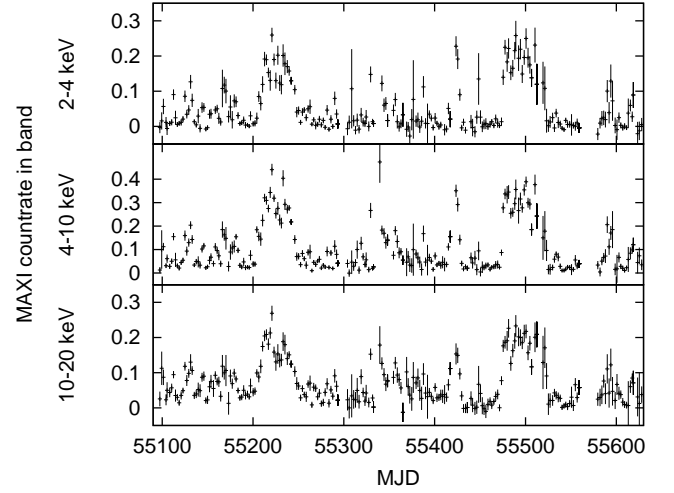}
\caption{MAXI lightcurve of Cen\,X-3 from MJD 55097 to 55630. The three energy bands 2--4\,keV, 4--10\,keV, and 10--20\,keV are plotted.}
\label{figure:4}
\end{figure}

\begin{figure} [!h] \centering
\includegraphics[viewport=0.0cm 0cm 17.8cm 25cm,scale=0.345,clip,angle=270]{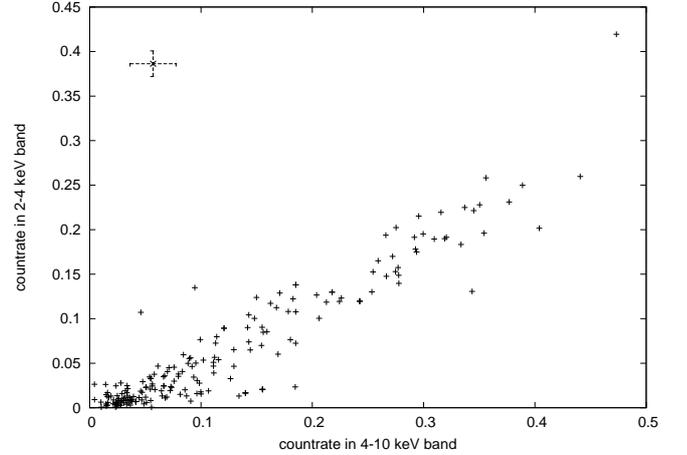}
\caption{MAXI data: Countrate in the 2--4\,keV band versus countrate in the 4--10\,keV band. The size of typical error bars is shown in the upper left of this figure.}
\label{figure:5}
\end{figure}

The long-term Cen\,X-3 lightcurve in all three energy bands of MAXI is shown in 
Figure\,\ref{figure:4}. As for the ASM data we plotted the fluxes in different
energy bands versus each other. Also with the MAXI data 
we could not confirm 
the presence of the two states reported for Cen\,X-3.
The two spectral modes are expected to appear best in a plot of the 2--4\,keV 
versus 4--10\,keV fluxes as those energy bands are closest to the bands ``B''
and ``C'' of ASM where the most clear separation in two spectral states is 
seen in \citet{2005A&A...442L..15P}. However, we could not find any 
indication of the two branches (see Fig.\,\ref{figure:5}).

\subsection{INTEGRAL results}

\begin{figure} [!h] \centering
\includegraphics[viewport=0.0cm 0cm 17.8cm 25cm,scale=0.345,clip,angle=270]{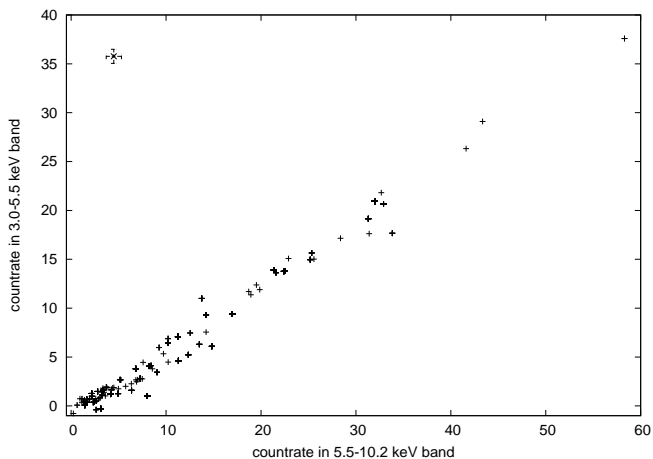}
\caption{INTEGRAL: Countrate in the 3.0--5.5\,keV band versus countrate in the 5.5--10.2\,keV band. The size of typical error bars is shown in the upper left of this figure.}
\label{figure:6}
\end{figure}

Following the procedure of \citet{2005A&A...442L..15P} we made countrate versus countrate plots for the selected JEM-X and ISGRI energy bands. As it follows from the results of \citet{2005A&A...442L..15P}, the separation of the two spectral states should appear best in a 3.0--5.5\,keV versus 5.5--10.2\,keV countrate plot of JEM-X. As for the MAXI data we could not find the presence of two spectral states (see Fig.\,\ref{figure:6}). Extending the energy bands up to higher energies, e.g. using ISGRI data, did not reveal any spectral separation of the source as well.

\section{Conclusions}

Triggered by the finding of \citet{2005A&A...442L..15P}, we performed a 
study of Cen\,X-3 using the data taken with two all sky monitors, 
RXTE/ASM and the MAXI as well as data taken from the JEM-X and ISGRI instruments of INTEGRAL. 
Using the same data and analysis procedures as \citet{2005A&A...442L..15P},
we could not find any spectral transitions either in the lightcurve or 
by plotting fluxes in different energy bands versus each other. 
The result does not change with addition of newest RXTE/ASM, MAXI or INTEGRAL data or by extending the analysis up to higher energy bands. 
Although \citet{2005A&A...442L..15P} ruled out possible instrumental effects 
on the basis of their analysis of the ASM data of some other 
sources for the same time interval as for Cen\,X-3, we suggest that systematic
and/or analysis effects are responsible for the previously 
reported appearance of the two spectral states. 
To clarify this issue we have contacted the ASM instrument team inquiring whether any substantial recalibration of ASM data took place after the work of \citet{2005A&A...442L..15P}. According to them, major calibrational changes occurred around MJD 51956 when telemetry modes switched from ASM "Position Histogram" to ASM Event mode. ASM camera SSC\,1 was mainly influenced by this change and a discontinuity around that time might have led to a change in the observed fluxes as reported in \citet{2005A&A...442L..15P}. However, the spectral discontinuity was seen in each ASM camera separately for Cen\,X-3, but not seen at all for a set of other pulsars. Additionally, the data software ran through major updates in April 2005 and 2007. This could have also led to the reported behavior if the data for Her\,X-1, Vela\,X-1, and SMC\,X-1 used by \citet{2005A&A...442L..15P} to check for possible instrumental effects were downloaded after the software update. The instrument team also generally claimed that although it is now difficult to check if the recalibration or data analysis software updates could lead to the reported effect, the regular improvement of the ASM calibration over time suggests that any later analysis is generally more reliable as the earlier one.

\begin{acknowledgements} This research has made use of the MAXI data provided by RIKEN, JAXA and the MAXI team, quick-look results provided by the ASM/RXTE team as well as data provided by the INTEGRAL Science Data Centre. We want to thank R. Remillard of the ASM/RXTE team for a detailed report of the history of ASM calibration and software updates. We also thank the referee for useful suggestions on improving the manuscript. This work has been partially funded by the DLR, grant 50 OR 1008, and by the Carl-Zeiss-Stiftung.\end{acknowledgements}

\bibliographystyle{aa}
\bibliography{articles}
\end{document}